\documentclass[reprint,superscriptaddress,amsmath,amssymb,aps,pra,prstab,floatfix]{revtex4-1}

\usepackage{graphicx}
\usepackage{dcolumn}
\usepackage{bm}
\usepackage{xcolor}

\usepackage{tikz}
\usepackage{pgfplots}
\usepgfplotslibrary{groupplots,units}
\usetikzlibrary{decorations.pathmorphing, spy}

\begin{document}


\title{Ramsey interferometry with arbitrary coherent-population-trapping pulse sequence}

\author{Ruihuan Fang}
\affiliation{Guangdong Provincial Key Laboratory of Quantum Metrology and Sensing $\&$ School of Physics and Astronomy, Sun Yat-Sen University (Zhuhai Campus), Zhuhai 519082, China}
\affiliation{State Key Laboratory of Optoelectronic Materials and Technologies, Sun Yat-Sen University (Guangzhou Campus), Guangzhou 510275, China}
\affiliation{College of Physics and Optoelectronic Engineering, Shenzhen University, Shenzhen 518060, China}

\author{Chengyin Han}
\affiliation{College of Physics and Optoelectronic Engineering, Shenzhen University, Shenzhen 518060, China}%

\author{Bo Lu}
\affiliation{College of Physics and Optoelectronic Engineering, Shenzhen University, Shenzhen 518060, China}%

\author{Jiahao Huang}
\email{hjiahao@mail2.sysu.edu.cn; eqjiahao@gmail.com}
\affiliation{Guangdong Provincial Key Laboratory of Quantum Metrology and Sensing $\&$ School of Physics and Astronomy, Sun Yat-Sen University (Zhuhai Campus), Zhuhai 519082, China}
\affiliation{College of Physics and Optoelectronic Engineering, Shenzhen University, Shenzhen 518060, China}%

\author{Chaohong Lee}
\email{chleecn@szu.edu.cn}
\affiliation{College of Physics and Optoelectronic Engineering, Shenzhen University, Shenzhen 518060, China}%

\date{\today}

\begin{abstract}
  Coherent population trapping (CPT) is a multi-level quantum coherence phenomenon of promising applications in atomic clocks and magnetometers.
  Particularly, multi-pulse CPT-Ramsey interferometry is a powerful tool for improving the performance of CPT atomic clocks.
  Most studies on multi-pulse CPT-Ramsey interferometry consider periodic pulse sequence and time-independent detuning.
  However, to further improve the accuracy and precision, one may modify the spectrum symmetry which involves pulse sequence with time-dependent detuning or phase shift.
  %
  %
  Here, we theoretically analyze the multi-pulse CPT-Ramsey interferometry under arbitrary pulse sequences of time-dependent detuning and obtain a general analytical formula.
  Using our formula, we analyze the popular CPT-Ramsey interferometry schemes such as two-pulse symmetric and antisymmetric spectroscopy, and multi-pulse symmetric and antisymmetric spectroscopy.
  Moreover, we quantitatively obtain the influences of pulse width, pulse period, pulse number, and Rabi frequency under periodic pulses.
  Our theoretical results can guide the experimental design to improve the performance of atomic clocks via multi-pulse CPT-Ramsey interferometry.
\end{abstract}

\maketitle


\section{Introduction}

Coherent population trapping (CPT) is a phenomenon of atoms trapped in a coherent state that does not interact with external laser fields.
Since the first observation of CPT spectrum~\cite{alzetta1976experimental}, it has been extensively utilized in various applications of quantum engineering and quantum sensing, such as all-optical manipulation~\cite{PhysRevLett.113.263602, PhysRevA.97.033838, PhysRevLett.115.093602, PhysRevLett.97.247401, PhysRevLett.116.043603, Ni231}, atomic cooling~\cite{ PhysRevLett.61.826}, atomic clocks~\cite{Vanier2005, Merimaa:03, PhysRevApplied.7.014018, PhysRevApplied.8.054001}, and atomic magnetometers~\cite{PhysRevLett.69.1360, Nagel_1998, doi:10.1063/1.1839274, Tripathi2019}.
Conventionally, CPT spectroscopy has the drawback of power broadening caused by strong CPT light power.
Using two CPT pulses to perform CPT-Ramsey interferometry can narrow the spectral linewidth and improve the signal-to-noise ratio (SNR)~\cite{PhysRevLett.94.193002,Merimaa:03, PhysRevA.67.065801}.
%
In this case, the linewidth of CPT-Ramsey spectrum can be narrower as the interval dark time between the two pulses increases~\cite{Merimaa:03, PhysRevA.67.065801}.
%
%
As the demands for higher measurement precision and accuracy grow, various techniques are developed to improve the spectrum SNR and resolution, as well as mitigate light shift.

The multi-pulse CPT-Ramsey interferometry has been developed in recent years~\cite{4126869, Yun_2012, Warren2018, Nicolas2018}.
The spectrum linewidth can be narrowed and meanwhile, the central peak can be identified due to the multi-pulse interference.
%
%
Understanding the mechanism of multi-pulse CPT-Ramsey interferometry is beneficial to designing suitable CPT pulse sequences for frequency measurement.
Some typical multi-pulse CPT-Ramsey schemes can be analytically analyzed.
For example, under multiple pulses with identical periods and duration, one can explain the multi-pulse CPT-Ramsey interference using a simple model based on the Fourier analysis of the CPT pulse sequence~\cite{Warren2018, PhysRevLett.116.043603}.
The Fourier analysis introduces a characteristic number $N_s$ as the spectrum will reach steady-state for large pulse number $N$~\cite{PhysRevLett.116.043603}.
The other analytical treatment is to compare the multi-pulse CPT-Ramsey interferometry to the Fabry-P\'erot resonator,
which is valid from periodic CPT pulse sequence~\cite{Nicolas2018} to arbitrary time-independent CPT pulse sequence~\cite{fang_temporal_2021}.

However, to further improve the performances one may need to modulate the pulse detuning or phase shift with time.
For example, the auto-balance technique uses a detuning change during the dark time to modify the symmetry of the spectrum to achieve real-time clock servo.
The frequency shift and phase jump have been applied to change the vertical symmetry of the spectrum and are used as the additional variables in auto-balanced CPT~\cite{abdel_hafiz_toward_2018,abdel_hafiz_symmetric_2018,yudin_generalized_2018}.
Moreover, for quantum lock-in amplifier~\cite{kotler_single-ion_2011}, one may even use the mixing between the alternating magnetic field and CPT pulse sequence.
The alternating magnetic field may also induce alternating detuning.
The modulation of frequency or phase has been applied in experimental CPT schemes and become a potential technique.
Thus the analytical analysis of multi-pulse CPT-Ramsey interferometry under time-dependent detuning and arbitrary pulse phase is of great importance and broad applications.

In this article, we study the multi-pulse CPT-Ramsey interferometry with arbitrary CPT pulse sequences.
In particular, when the single-photon detuning and Rabi frequencies are relatively small compared to the decay rate of the excited state,
we obtain an analytical formula to analyze various CPT-Ramsey interferometry scenarios, such as the conventional CPT-Ramsey interferometry, CPT-Ramsey interferometry with frequency shift or phase jump, multi-pulse CPT-Ramsey interferometry, and multi-pulse CPT-Ramsey interferometry with frequency shift.
For the periodic CPT pulse sequence~\cite{Yun_2012}, we can analytically study the influences of pulse length, pulse strength, pulse interval, and pulse number on the spectrum linewidth.
The formula we derived is a general solution that is valid for most situations with arbitrary CPT pulse sequences.

\section{\label{sec:citeref}CPT in three-level $\Lambda$ system}

\subsection{Model}

\begin{figure}
  \centering
  \includegraphics{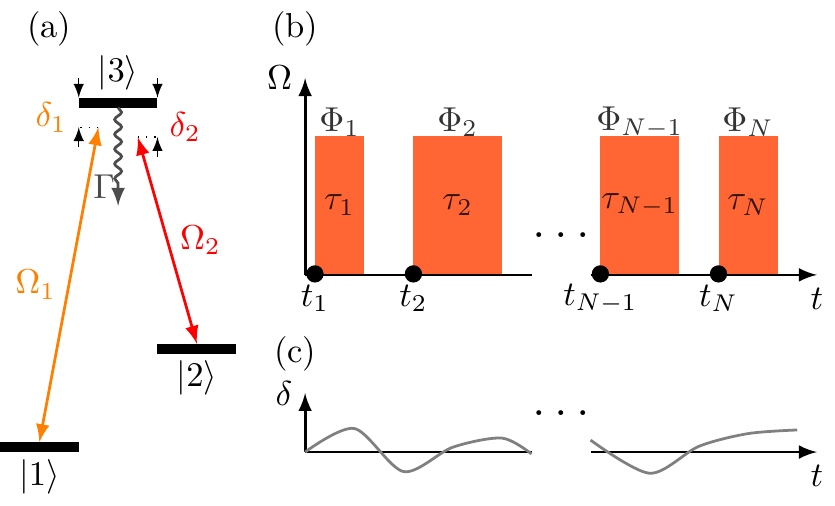}
    \caption{\textbf{The three-level $\Lambda$ system and the timing sequence of CPT.} (a) Bi-chromatic light with the Rabi frequencies $\Omega_1$ and $\Omega_2$ couple the two ground states $|1\rangle$ and $|2\rangle$ to the excited state $|3\rangle$.
    The $\delta_1$ and $\delta_2$ are the frequency detunings from the $|1\rangle$ and $|2\rangle$ to $|3\rangle$.
    The decay rate of excited state $|3\rangle$ is $\Gamma$.
    (b) The orange pulses are CPT pulses with the Rabi frequency $\Omega$. The $\tau_i$ is the pulse length, $t_i$ is the start point, and $\Phi_i$ is the phase of $i$-th CPT pulse.
    (c) The $\delta$ versus time $t$. The $\delta$ can be time-dependent randomly.
    }
    \label{Fig:ThreeLevel}
  \end{figure}

CPT can usually be achieved in a three-level $\Lambda$ structure as shown in Fig.~\ref{Fig:ThreeLevel}.
The bi-chromatic field with Rabi frequencies $\Omega_1$ and $\Omega_2$ couple the two ground states $|1\rangle$ and $|2\rangle$ to the excited state $|3\rangle$.
Here, $\delta_1$ and $\delta_2$ are the corresponding detunings, and $\Gamma$ is the decay rate of the excited state $|3\rangle$.
Under the rotating wave approximation, the Hamiltonian in the interaction picture reads~\cite{Shahriar2014}
\begin{equation}
  \begin{aligned}
  \hat{H}_I=&\hbar\left(\delta_1|1\rangle\langle 1|+\delta_2|2\rangle\langle 2|\right)\\
  &+\hbar\left(\frac{\Omega_1}{2}|1\rangle\langle 3| + \frac{\Omega_2}{2}|2\rangle\langle 3| + h.c.\right).
  \end{aligned}
\end{equation}
According to the Lindblad equation, the time-evolution of the density matrix $\rho$ obeys
\begin{equation}
  \frac{d\rho}{d t} = -\frac{\mathrm{i}}{\hbar}\left[\hat{H}_I, \rho\right] + \frac{\Gamma}{2} \sum_{j=1}^{2} \left[\hat{L}_j\rho\hat{L}_j^\dagger
  -\frac{1}{2}\{\hat{L}_j^\dagger\hat{L}_j, \rho\}\right].
  \label{Eq:Lindblad}
\end{equation}
Here, $\hat{L}_j=|j\rangle\langle 3|$ and their Hermite conjugate $\hat{L}_j^\dagger$ are the Lindblad operators.
In most cases, the Rabi frequencies of monochromatic light are equal, i.e., $|\Omega_1|=|\Omega_2|=\Omega$.
Then the Rabi frequencies can be expressed as $\Omega_1=\Omega\mathrm{e}^{\mathrm{i}\phi_1}$ and $\Omega_2=\Omega\mathrm{e}^{\mathrm{i}\phi_2}$, with $\phi_1$ and $\phi_2$ are the phases of monochromatic light.
In this case, Eq.~\eqref{Eq:Lindblad} can be written as follows,
\begin{equation}
  \begin{cases}
    \frac{d}{dt}\rho_{11} = \frac{\Gamma}{2}\rho_{33} - \mathrm{i} \left(\frac{\Omega\mathrm{e}^{-\mathrm{i}\phi_1}}{2}\rho_{13}^* - \frac{\Omega\mathrm{e}^{\mathrm{i}\phi_1}}{2}\rho_{13}\right)\\
    \frac{d}{dt}\rho_{12} = -\mathrm{i}\frac{\Omega\mathrm{e}^{-\mathrm{i}\phi_1}}{2}\rho_{23}^* + \mathrm{i}\frac{\Omega\mathrm{e}^{\mathrm{i}\phi_2}}{2}\rho_{13} - \mathrm{i}\left(\delta_1-\delta_2\right)\rho_{12}\\
    \frac{d}{dt}\rho_{13} = -\frac{\Gamma}{2}\rho_{13} + \mathrm{i}\frac{\Omega\mathrm{e}^{-\mathrm{i}\phi_1}}{2}\rho_{11} - \mathrm{i}\frac{\Omega\mathrm{e}^{-\mathrm{i}\phi_1}}{2}\rho_{33} \\
    ~~~~~~~~~~~~~~+ \mathrm{i}\frac{\Omega\mathrm{e}^{-\mathrm{i}\phi_2}}{2}\rho_{12} - \mathrm{i}\delta_1\rho_{13}\\
    \frac{d}{dt}\rho_{12}^* = \mathrm{i}\frac{\Omega\mathrm{e}^{\mathrm{i}\phi_1}}{2}\rho_{23} - \mathrm{i}\frac{\Omega\mathrm{e}^{-\mathrm{i}\phi_2}}{2}\rho_{13}^* + \mathrm{i}\left(\delta_1-\delta_2\right)\rho_{12}^*\\
    \frac{d}{dt}\rho_{22} = \frac{\Gamma}{2}\rho_{33} - \mathrm{i}\left(\frac{\Omega\mathrm{e}^{-\mathrm{i}\phi_2}}{2}\rho_{23}^* - \frac{\Omega\mathrm{e}^{\mathrm{i}\phi_2}}{2}\rho_{23}\right)\\
    \frac{d}{dt}\rho_{23} = -\frac{\Gamma}{2}\rho_{23} + \frac{\mathrm{i}\Omega\mathrm{e}^{-\mathrm{i}\phi_1}}{2}\rho_{12}^* + \frac{\mathrm{i}\Omega\mathrm{e}^{-\mathrm{i}\phi_2}}{2}\rho_{22} \\
    ~~~~~~~~~~~~~~~- \frac{\mathrm{i}\Omega\mathrm{e}^{-\mathrm{i}\phi_2}}{2}\rho_{33}-\mathrm{i}\delta_{2}\rho_{23}\\
    \frac{d}{dt}\rho_{13}^* = -\frac{\Gamma}{2}\rho_{13}^* - \mathrm{i}\frac{\Omega\mathrm{e}^{\mathrm{i}\phi_1}}{2}\rho_{11}+\mathrm{i}\frac{\Omega\mathrm{e}^{\mathrm{i}\phi_1}}{2}\rho_{33}\\
    ~~~~~~~~~~~~~~-\mathrm{i}\frac{\Omega\mathrm{e}^{\mathrm{i}\phi_2}}{2}\rho_{12}^* + \mathrm{i}\delta_1\rho_{13}^*\\
    \frac{d}{dt}\rho_{23}^* = -\frac{\Gamma}{2}\rho_{23}^* - \frac{\mathrm{i}\Omega\mathrm{e}^{\mathrm{i}\phi_1}}{2}\rho_{12} - \frac{\mathrm{i}\Omega\mathrm{e}^{\mathrm{i}\phi_2}}{2}\rho_{22} \\
    ~~~~~~~~~~~~~~+ \frac{\mathrm{i}\Omega\mathrm{e}^{\mathrm{i}\phi_2}}{2}\rho_{33} + \mathrm{i}\delta_2\rho_{23}^*\\
    \frac{d}{dt}\rho_{33} = -\Gamma\rho_{33} + \frac{\mathrm{i}\Omega\mathrm{e}^{-\mathrm{i}\phi_1}}{2}\rho_{13}^* - \frac{\mathrm{i}\Omega\mathrm{e}^{\mathrm{i}\phi_1}}{2}\rho_{13} \\
    ~~~~~~~~~~~~~~+ \frac{\mathrm{i}\Omega\mathrm{e}^{-\mathrm{i}\phi_2}}{2}\rho_{23}^* - \frac{\mathrm{i}\Omega\mathrm{e}^{\mathrm{i}\phi_2}}{2}\rho_{23},
  \end{cases}
  \label{Eq:master}
\end{equation}
where $\rho_{ij}=\langle i|\rho|j\rangle$. In the following, we solve these equations analytically.
\subsection{Analytical formula of ground-state coherence}
Under the condition of the Rabi frequency is far smaller than the decay rate $\Omega\ll\Gamma$, one can find $\frac{d}{d t}\rho_{33}\ll \Gamma\rho_{33}$, and we can get
\begin{equation}
  \rho_{33} = \frac{1}{\Gamma} \left(\frac{\mathrm{i}\Omega\mathrm{e}^{-\mathrm{i}\phi_1}}{2}\rho_{13}^* - \frac{\mathrm{i}\Omega\mathrm{e}^{\mathrm{i}\phi_1}}{2}\rho_{13} + \frac{\mathrm{i}\Omega\mathrm{e}^{-\mathrm{i}\phi_2}}{2}\rho_{23}^* - \frac{\mathrm{i}\Omega\mathrm{e}^{\mathrm{i}\phi_2}}{2}\rho_{23}\right).
\label{Eq:rho33}
\end{equation}
Usually, the CPT works in the situation of near resonance $\delta_{1,2}\ll \Gamma$.
Due to the large decay rate of the excited state, the population in the excited state can be ignored (compared with ground state populations), i.e.,  $\rho_{33}\ll\rho_{11}$ and $\rho_{33}\ll\rho_{22}$.
Thus we can obtain
\begin{equation}
  \rho_{13} = \frac{1}{\Gamma}\left(\mathrm{i}\Omega\mathrm{e}^{-\mathrm{i}\phi_1}\rho_{11} + \mathrm{i}\Omega\mathrm{e}^{-\mathrm{i}\phi_2}\rho_{12}\right),
  \label{Eq:rho13}
\end{equation}
\begin{equation}
  \rho_{23} = \frac{1}{\Gamma}\left(\mathrm{i}\Omega\mathrm{e}^{-\mathrm{i}\phi_1}\rho_{12}^* + \mathrm{i}\Omega\mathrm{e}^{-\mathrm{i}\phi_2}\rho_{22}\right),
  \label{Eq:rho23}
\end{equation}
\begin{equation}
  \rho_{13}^* = \frac{1}{\Gamma}\left(- \mathrm{i}\Omega\mathrm{e}^{\mathrm{i}\phi_1}\rho_{11}-\mathrm{i}\Omega\mathrm{e}^{\mathrm{i}\phi_2}\rho_{12}^*\right),
  \label{Eq:rho13*}
\end{equation}
\begin{equation}
  \rho_{23}^* = \frac{1}{\Gamma}\left(-\mathrm{i}\Omega\mathrm{e}^{\mathrm{i}\phi_1}\rho_{12} - \mathrm{i}\Omega\mathrm{e}^{\mathrm{i}\phi_2}\rho_{22}\right).
  \label{Eq:rho23*}
\end{equation}
In a CPT process, the population is mostly in the ground states that $\rho_{11}+\rho_{22}\approx 1$.
Substituting Eqs.~\eqref{Eq:rho13}-~\eqref{Eq:rho23*} into Eqs.~\eqref{Eq:master} and~\eqref{Eq:rho33}, we obtain
\begin{equation}
  \rho_{33} = \frac{\Omega^2}{\Gamma^2}\left(1+2\Re(\rho_{12}\mathrm{e}^{\mathrm{i}\phi(t)})\right),
  \label{Eq:rho33_2}
\end{equation}
and
\begin{equation}
\frac{d}{dt}\rho_{12}=-\frac{\Omega^2}{2\Gamma}\mathrm{e}^{-\mathrm{i}\phi(t)} - \left(\frac{\Omega^2}{\Gamma}+ \mathrm{i}\delta\right)\rho_{12}.
\label{Eq:rho12}
\end{equation}
Here, $\delta=\delta_1-\delta_2$ is the two-photon detuning, and $\phi(t)=\phi_1-\phi_2$ is the phase of monochromatic light.
%
%
Eq.~\eqref{Eq:rho12} can be analytically solved under a train of CPT pulses.

We can construct the response of Eq.~\eqref{Eq:rho12} to an impulse taking place at $t_0$, which is the Green's function~\cite{berberan-santos_greens_2010}
\begin{equation}
  G(t, t_0) = H(t-t_0)\exp\left(-\int_{t_0}^t \left(\frac{\Omega^2(u)}{\Gamma}+\mathrm{i}\delta(u)\right)du\right),
\end{equation}
where $H(x)$ is Heaviside's function.
If the density matrix starts from a mixture state $\rho=|1\rangle\langle 1|+|2\rangle\langle 2|$.
%
%
Take the initial value $\rho_{12}(0)=0$, we have
\begin{equation}
  \rho_{12}(t, \Omega)=-\int_{0}^{t}\frac{\Omega^2(t^\prime)}{2\Gamma}\mathrm{e}^{-\mathrm{i}\phi(t^\prime)}G(t,t^\prime)dt^\prime
  \label{Eq:solution}
\end{equation}
being a function of $\delta(t)$ and $\Omega(t)$.
We analyze Eq.~\eqref{Eq:solution} within the context of a general CPT pulse sequence, as shown in Figs.~\ref{Fig:ThreeLevel}(b,c).
The orange pulses represent the CPT pulses with Rabi frequency $\Omega$, variable pulse duration $\tau_i$, and phase $\Phi_i$.
The gray solid line represents the time-dependent detuning $\delta(t)$.
As a result, Eq.~\eqref{Eq:solution} can be expressed as
\begin{equation}
  \begin{aligned}
    \rho_{12}(t, \Omega)=&-\frac{\Omega^2}{2\Gamma}\sum_{l=1}^{N}\left(\prod_{k=l+1}^{N}\exp\left(-\frac{\Omega^2}{\Gamma}\tau_k\right)\right)\\
    &\times\exp\left(- \int_{t_l+\tau_l}^{t}\mathrm{i}\delta(u) du \right)\mathrm{e}^{-\mathrm{i}\Phi_l}\\
    &\times \int_{t_l}^{t_l+\tau_l}\exp\left(- \int_{t^\prime}^{t_l+\tau_l}\left(\frac{\Omega^2}{\Gamma}+\mathrm{i}\delta(u)\right)du\right)dt^\prime.
  \end{aligned}
  \label{Eq:analytical}
\end{equation}
This is the general formula of ground-state coherence.
Eq.~\eqref{Eq:analytical} can be used to describe most cases of pulse sequences including the two-pulse CPT-Ramsey interferometry and multi-pulse CPT-Ramsey interferometry under fixed or time-dependent frequency detuning and phase.
However, the absence of single-photon detunings $\delta_1$ and $\delta_2$ in the derivation of Eqs.~\eqref{Eq:rho13}-\eqref{Eq:rho23*} means that the influences of light shift induced by the excited state are not taken into account.

According to Eq.~\eqref{Eq:rho33_2}, if the phases of each CPT pulse are identical, the phase value does not affect the observation $\rho_{33}$.
This is because a global phase of the Rabi frequencies can be gauged into the state without changing the density matrix if we select the initial mixture state without non-diagonal terms.
That means the magnitude of Rabi frequencies can be real if we select proper initial phases of $|1\rangle$ and $|2\rangle$.
Generally, we gauge the phase of the last CPT pulse $\Phi_N$ into zero such that the phase of Eq.~\eqref{Eq:rho33_2} can be eliminated, thus
\begin{equation}
  \rho_{33}=\frac{\Omega^2}{\Gamma^2}\left(1+2\Re\left(\rho_{12}\right)\right).
\end{equation}
And if the phases of CPT pulses are constant, all the phases can be gauged as zero.

If the detuning $\delta(t)$ is fixed as $\delta$ during the CPT pulses and varies with time $t$ during the dark period, Eq.~\eqref{Eq:analytical} can be written in the form of
\begin{equation}
  \rho_{12} = f\left(\delta\right) Q,
  \label{Eq:fq}
\end{equation}
where
\begin{equation}
  f(\delta)=-\frac{\Omega^2}{2\Gamma\left(\frac{\Omega^2}{\Gamma}+\mathrm{i}\delta\right)},
  \label{Eq:fdelta}
\end{equation}
is the slow variant envelope and $Q$ is the multi-pulse CPT-Ramsey interference term,
\begin{equation}
  \begin{aligned}
    Q = \sum_{l=1}^{N}&\left(\prod_{k=l+1}^{N}\mathrm{e}^{-\frac{\Omega^2}{\Gamma}\tau_k}\right)\exp\left(- \int_{t_l+\tau_l}^{t}\mathrm{i}\delta(u) du \right)\mathrm{e}^{-\mathrm{i}\Phi_l}\\
    &\times \left(1-\exp\left(-\left(\frac{\Omega^2}{\Gamma}+\mathrm{i}\delta\right)\tau_l\right)\right).
    \end{aligned}
  \label{Eq:OriginalQ}
\end{equation}

\section{Applications in CPT-Ramsey Spectroscopy}

Our analytical formula Eq.~\eqref{Eq:analytical} can be applied in various experimental CPT-Ramsey scenarios.
In the following, we show its applications in conventional two-pulse CPT-Ramsey spectroscopy, anti-symmetric two-pulse CPT-Ramsey spectroscopy with frequency shift, and multi-pulse CPT-Ramsey spectroscopy.

\subsection{Two-pulse sequence}\label{subA}

In most cases, the CPT-Ramsey interferometry is implemented with two pulses, as shown in Fig.~\ref{Fig:Ramsey}~(a).
In this scheme, the CPT pulse sequence contains a preparation pulse with duration $\tau_1$ and a detection pulse with duration $\tau_2$, which are separated by a free evolution with dark time $T$~\cite{zanon-willette_high_2005}.
Using Eq.~\eqref{Eq:analytical} with pulse number $N=2$, we can easily get the corresponding analytical results, which can be used as a benchmark example.
Our analytical results can also be used for optimizing the CPT pulse sequence as we need.

\subsubsection{Conventional Two-pulse CPT-Ramsey Spectroscopy}

Considering the simple case, in which the detunings are time-independent and the phases equal zero. Then, Eq.~\eqref{Eq:analytical} can be simplified as
\begin{equation}
  \begin{aligned}
  \rho_{12}(t, \delta, \Omega)=&-\frac{\Omega^2}{2\Gamma\left(\frac{\Omega^2}{\Gamma}+\mathrm{i}\delta\right)}\sum_{l=1}^{N}\left(\prod_{k=l+1}^{N}\exp\left(-\frac{\Omega^2}{\Gamma}\tau_{k}\right)\right)\\
  &\times \exp\left(-\mathrm{i}\delta\left(t-t_l-\tau_l\right)\right)\\
  &\times \left(1-\exp\left(-\left(\frac{\Omega^2}{\Gamma}+\mathrm{i}\delta\right)\tau_l\right)\right).
  \end{aligned}
  \label{Eq:constantDetuning}
\end{equation}
\begin{figure}
  \centering
  \includegraphics{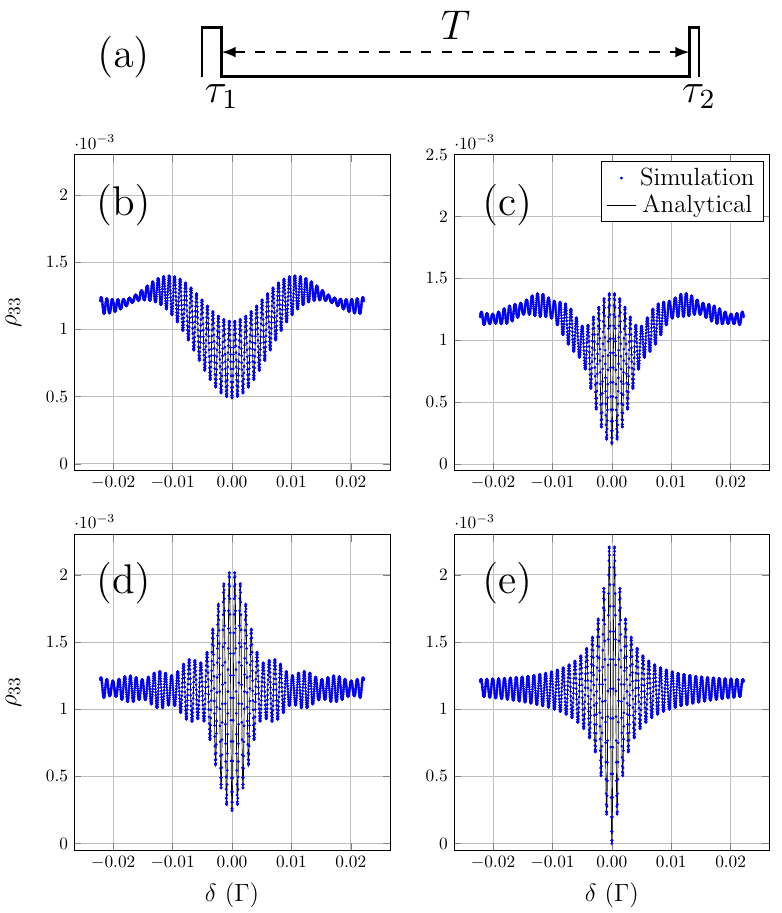}
  \caption{\textbf{The two-pulse CPT-Ramsey spectrum.}
  (a) The timing sequence of conventional CPT-Ramsey includes the preparation pulse $\tau_1$, the detection pulse $\tau_2$, and the dark time $T$.
  The Rabi frequency is $\Omega=0.035\Gamma$.
  The CPT-Ramsey spectrum of both simulation(blue dots) and analytical(black solid line) are obtained with dark time $T=1000\times\frac{2\pi}{\Gamma}$ as well as the pulse sequences of
  (b) $\tau_1=60 \times \frac{2\pi}{\Gamma}$ and $\tau_2=60 \times \frac{2\pi}{\Gamma}$, (c) $\tau_1=200 \times \frac{2\pi}{\Gamma}$ and $\tau_2=60 \times \frac{2\pi}{\Gamma}$, (d) $\tau_1=200 \times \frac{2\pi}{\Gamma}$ and $\tau_2=10 \times \frac{2\pi}{\Gamma}$, (e)$\tau_1=1000 \times \frac{2\pi}{\Gamma}$ and $\tau_2=10 \times \frac{2\pi}{\Gamma}$.
  %
  }
  \label{Fig:Ramsey}
\end{figure}
For a two-pulse CPT-Ramsey interferometry, the CPT-Ramsey interference term reads
\begin{equation}
  Q = Q^T_1 + Q^T_2,
  \label{Eq:Q1}
\end{equation}
where
\begin{equation*}
  \begin{aligned}
    Q_1^T =& \exp\left(-\frac{\Omega^2}{\Gamma}\tau_{2}\right)\left(1-\exp\left(-\left(\frac{\Omega^2}{\Gamma}+\mathrm{i}\delta\right)\tau_1\right)\right)\\
    &\times \exp\left(-\mathrm{i}\delta\left(T+\tau_2\right)\right),\\
  \end{aligned}
\end{equation*}
and
\begin{equation*}
  Q_2^T = \left(1-\exp\left(-\left(\frac{\Omega^2}{\Gamma}+\mathrm{i}\delta\right)\tau_2\right)\right).
\end{equation*}
The conventional CPT-Ramsey fringe is mainly dominated by $Q^T_1$ in Eq.~\eqref{Eq:Q1}.
It increases with the preparation pulse duration $\tau_1$ and decreases with the detection pulse duration $\tau_2$.
$Q^T_2$ in Eq.~\eqref{Eq:Q1} increases with the detection pulse $\tau_2$, which contributes a trend of slow variance, resulting in the vertical asymmetry~\cite{zanon-willette_high_2005}.

As a benchmark example, we examine the conventional CPT-Ramsey pulse sequence consisting of a preparation pulse of duration $\tau_1$, a detection pulse of duration $\tau_2$, and the pulse interval $T=1000\times\frac{2\pi}{\Gamma}$, as illustrated in Fig.~\ref{Fig:Ramsey}~(a).
The Rabi frequency of the CPT pulse is $\Omega=0.035\Gamma$.
For a detection pulse duration of $\tau_2=60\times\frac{2\pi}{\Gamma}$, if the preparation pulse duration is short, $\tau_1=60\times\frac{2\pi}{\Gamma}$, the CPT-Ramsey spectrum contrast is low, as shown in Fig.~\ref{Fig:Ramsey}~(b).
The black solid line is the analytical result of Eq.~\eqref{Eq:Q1} and blue dots are the numerical result of Lindblad equations.
As the preparation pulse duration $\tau_1$ increases to $\tau_1=200\times\frac{2\pi}{\Gamma}$, the $Q^T_1$ of Eq.~\eqref{Eq:Q1} grows, improving the contrast of the CPT-Ramsey fringe, see Fig.~\ref{Fig:Ramsey}~(c).
Meanwhile, reducing the detection pulse duration to $\tau_2=10\times\frac{2\pi}{\Gamma}$ further improves the contrast of the CPT-Ramsey fringe and the spectrum becomes vertically symmetric, as shown in Fig.~\ref{Fig:Ramsey}~(d).
When the preparation pulse duration $\tau_1=1000\times\frac{2\pi}{\Gamma}$ is sufficient, the spectrum will reach a saturation point, see Fig.~\ref{Fig:Ramsey}~(e).
As the period in frequency space of the $Q^T_1$ is $\frac{2\pi}{T+\tau_2}$.
The linewidth of the CPT-Ramsey spectrum satisfies $\Delta \nu = \frac{1}{2(T+\tau_2)}$.
Thus, the CPT-Ramsey with a long dark time $T$ can narrow the spectrum linewidth, which is consistent with our common knowledge.
These results indicate that our analytical formula is valid for conventional two-pulse CPT-Ramsey scenarios.

\subsubsection{Anti-symmetric Two-pulse CPT-Ramsey Spectroscopy with Frequency shift}\label{Sec:frequencyJump}

Usually, we need to perform two CPT-Ramsey interferometry with different frequencies and compare their differences to obtain an antisymmetric error signal for clock locking~\cite{Yun_2012}.
\begin{figure}
  \includegraphics{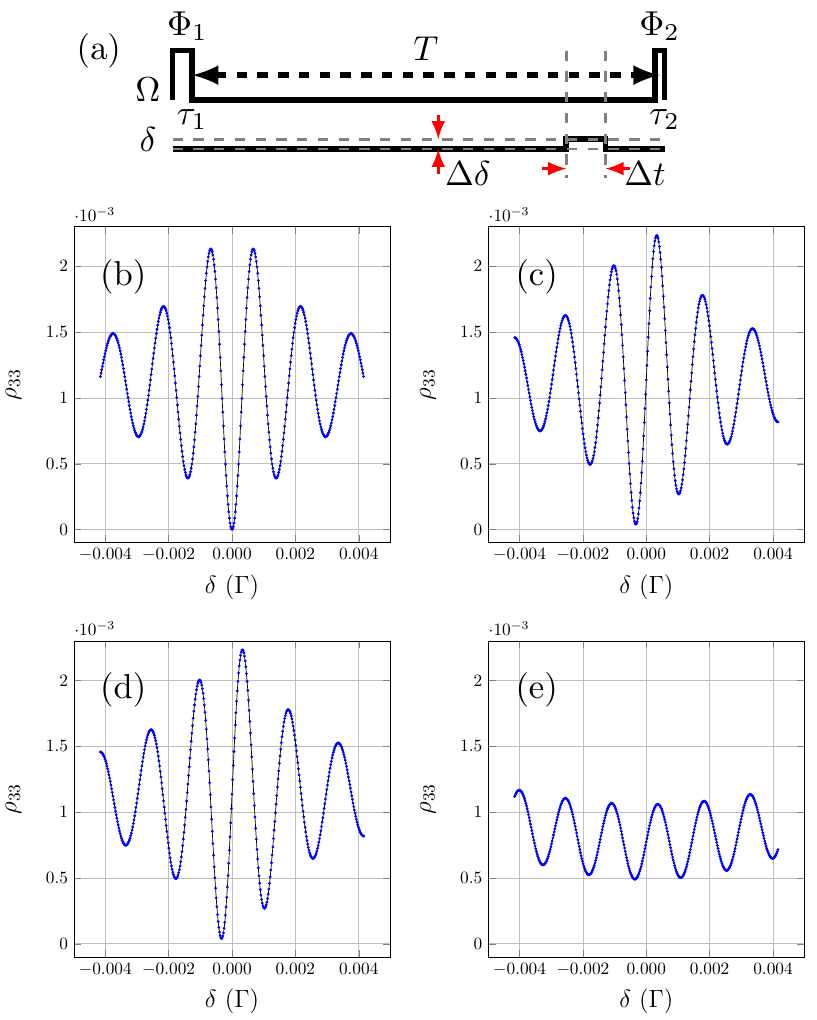}
  \caption{\textbf{Two-pulse CPT-Ramsey interferometry with frequency and phase shifts.}
  (a) The timing sequence of Rabi frequency $\Omega$ and detuning $\delta$ of two-pulse CPT-Ramsey sequences with time $T=600\times\frac{2\pi}{\Gamma}$. The amplitude of Rabi frequency is $0.035\Gamma$, the $\tau_1$ and $\tau_2$ are preparation and detection durations respectively,
  the $\Delta \delta$ and $\Delta t=100\times\frac{2\pi}{\Gamma}$ are the magnitude and duration of frequency shift.
  The $\Phi_1$ and $\Phi_2$ are the phases of preparation and detection pulses, respectively.
  The CPT-Ramsey spectra of numerical simulation (blue dots) and analytical formula (black solid line) with $\tau_1=1000\times\frac{2\pi}{\Gamma}$, $\tau_2=10\times\frac{2\pi}{\Gamma}$ as well as
  (b) zero frequency shift $\Delta\delta=0$ and $\Phi_1=\Phi_2$,
  (c) frequency shift $\Delta\delta=\frac{\pi/2}{\Delta t}$ and $\Phi_1=\Phi_2$,
  (d) zero frequency shift $\Delta\delta=0$, $\Phi_1=0$ and $\Phi_2=-\frac{\pi}{2}$.
  (e) The CPT-Ramsey spectra of numerical simulation (blue dots) and analytical formula (black solid line) with $\tau_1=60\times\frac{2\pi}{\Gamma}$, $\tau_2=60\times\frac{2\pi}{\Gamma}$ as well as zero frequency shift $\Delta\delta=0$, $\Phi_1=0$ and $\Phi_2=-\frac{\pi}{2}$.}
  \label{Fig:FrequencyShift}
\end{figure}
To achieve real-time antisymmetric spectra, we can directly apply a frequency shift of $\Delta\delta$ during the dark time $T$ or implement a change in phase $\Delta\Phi=\Phi_2-\Phi_1$ for the detection pulse based on conventional CPT-Ramsey interferometry, as illustrated in Fig.~\ref{Fig:FrequencyShift}~(a).
Here, $\Phi_1$ and $\Phi_2$ are the phases of the preparation pulse and detection pulse, respectively.
According to Eq.~\eqref{Eq:analytical}, the two-pulse CPT-Ramsey interference term becomes
\begin{equation}
  Q = Q^{T\prime}_1 + Q^{T}_2,
  \label{Eq:Q11}
\end{equation}
where
\begin{equation}
    Q^{T\prime}_1 =Q^{T}_1\exp\left(-\mathrm{i}\Delta\delta \Delta t + \mathrm{i}\Delta\Phi\right).
\end{equation}
As an example, we set $\tau_1=1000\times\frac{2\pi}{\Gamma}$, $\tau_2=10\times\frac{2\pi}{\Gamma}$, and $T=1000\times\frac{2\pi}{\Gamma}$.
When the two phases are equal, i.e., $\Phi_1=\Phi_2$ and $\Delta \Phi=0$, then it reduces to the conventional CPT-Ramsey interferometry as shown in Fig.~\ref{Fig:FrequencyShift}~(b).
While the spectra will become horizontal antisymmetric if the frequency shift and phase shift satisfy $\Delta\delta\Delta t-\Delta\Phi=\frac{\pi}{2}$, as shown in Fig.~\ref{Fig:FrequencyShift}~(c) and (d).
Then, the real-time processing of error signals will be obtained.
In auto-balance CPT-Ramsey interferometry~\cite{abdel_hafiz_symmetric_2018}, the $\Delta\Phi$ or $\Delta\delta$ can be used as the additional parameters to compensate for the light shift.
However, if $\tau_1=60\times\frac{2\pi}{\Gamma}$ is short and $\tau_2=60\times\frac{2\pi}{\Gamma}$ is substantial, the error signals will not be horizontally antisymmetric as $Q^T_2$ is considerable compared to $Q^{T\prime}_1$.
It means that if the detection pulse duration is comparable with the preparation pulse duration, the spectrum will become horizontally asymmetric, as shown in Fig.~\ref{Fig:FrequencyShift}~(e).
The black solid lines are the analytical result of Eq.~\eqref{Eq:Q11} and the blue dots are the corresponding numerical results.
All the results match perfectly.

\subsection{Multi-pulse sequence}

In Sec.~\ref{subA}, we analyzed conventional and frequency-shifted two-pulse CPT-Ramsey spectra using our analytical formula.
In this section, we will analyze multi-pulse CPT-Ramsey interferometry.
With a multi-pulse sequence, the central peak becomes obvious due to constructive interference while the neighboring peaks are suppressed through destructive interference.
The multi-pulse CPT-Ramsey interferometry makes the central peak easy to be identified and makes the signal-to-noise ratio (SNR) better.
Thus, the multi-pulse sequence is useful for developing practical quantum sensors, such as atomic clocks.
However, for practical applications, the multi-pulse CPT-Ramsey interferometry involves multiple pulses which need to be sophisticatedly tuned.
Our analytical formula provides a simple way to analyze and optimize the pulse sequence as desired.
We consider that the multi-pulse CPT Ramsey interferometry starts with a preparation pulse $\tau_1$ to prepare the dark state, followed by $N$ pulses of duration $\tau$ with pulse interval $T$, as shown in Fig.~\ref{Fig:multi-pulse}~(a).
By using our analytical formula, below we analyze the roles of preparation pulse and periodic pulse sequence and provide an example to achieve the anti-symmetric spectrum with frequency shift.

\subsubsection{The influence of preparation pulse}
\begin{figure}
  \includegraphics{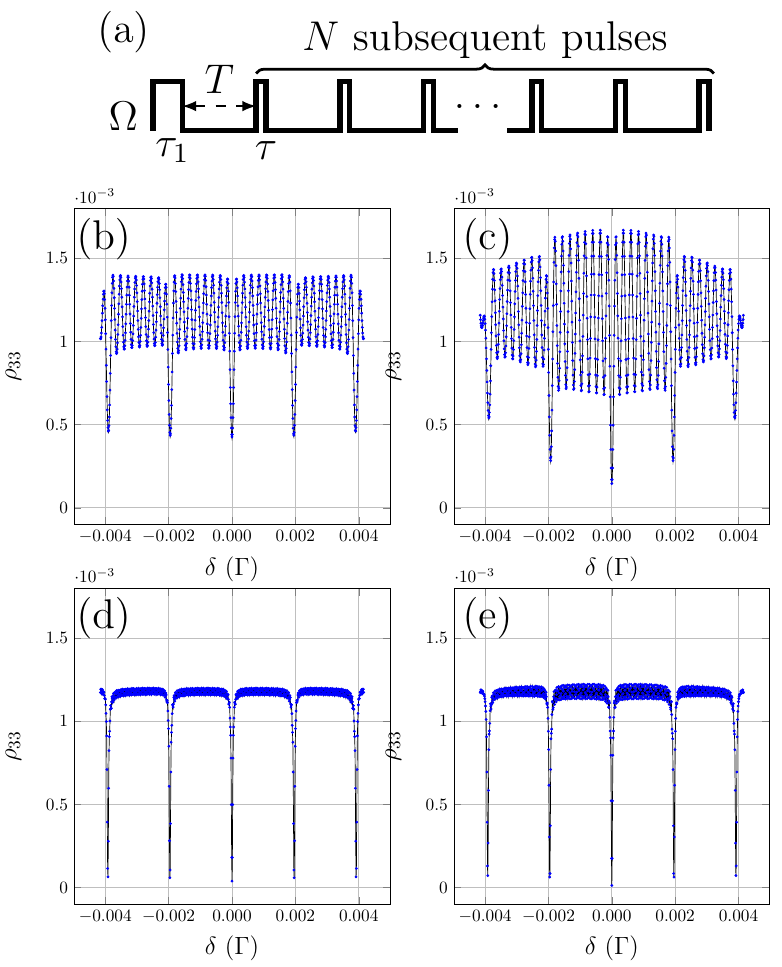}
  \caption{\textbf{Multi-pulse CPT-Ramsey interferometry.}
  (a) The timing sequence of Rabi frequency $\Omega$ of multi-pulse CPT-Ramsey sequences.
  Here, $\tau_1$ is the preparation pulse,$T=500\times\frac{2\pi}{\Gamma}$ is the dark time, $\tau=10\times\frac{2\pi}{\Gamma}$ is the pulse duration and $N$ is the pulse number.
  The multi-pulse CPT-Ramsey spectra with (b) $\tau_1=60\times\frac{2\pi}{\Gamma}$ and $N=8$, (c) $\tau_1=200\times\frac{2\pi}{\Gamma}$ and $N=8$, (d) $\tau_1=60\times\frac{2\pi}{\Gamma}$ and $N=40$,
  (e) $\tau_1=200\times\frac{2\pi}{\Gamma}$ and $N=40$.}
  \label{Fig:multi-pulse}
\end{figure}
We consider the phases of all CPT pulses are identical and so that we can set the phases as zero.
According to Eq.~\eqref{Eq:rho12}, the interference term
\begin{equation}
  Q=Q^M_1+Q^M_2,
\end{equation}
includes two parts,
\begin{equation*}
  \begin{aligned}
  Q^M_1 =& \exp\left(-\frac{\Omega^2}{\Gamma}N\tau\right)\exp\left(-\mathrm{i}\delta N\left(T+\tau\right)\right)\\
  &\times \left(1-\exp\left(-\left(\frac{\Omega^2}{\Gamma}+\mathrm{i}\delta\right)\tau_1\right)\right),
  \end{aligned}
\end{equation*}
and
\begin{equation*}
  \begin{aligned}
    Q^M_2 = &\sum_{i=1}^{N}\exp\left(-\frac{\Omega^2}{\Gamma}(N-i)\tau\right) \exp\left(-\mathrm{i}\delta (N-i)\left(T+\tau\right)\right)\\
    &\times \left(1-\exp\left(-\left(\frac{\Omega^2}{\Gamma}+\mathrm{i}\delta\right)\tau\right)\right).
  \end{aligned}
\end{equation*}
$Q^M_1$ is the fast oscillating term versus $\delta$.
Obviously, the preparation pulse duration $\tau_1$ only affects $Q^M_1$ and $Q^M_1$ increases with $\tau_1$.
%
%
Therefore the amplitude of the spectrum increases with the preparation pulse duration.
As shown in Fig.~\ref{Fig:multi-pulse}~(b) and (c), the spectra of multi-pulse CPT-Ramsey interferometry with longer preparation pulse duration $\tau_1=200\times\frac{2\pi}{\Gamma}$ has a higher amplitude of the peaks than that with shorter preparation pulse $\tau_1=60\times\frac{2\pi}{\Gamma}$.

However, the subsequent CPT pulses also affect $Q^M_1$.
The more or longer subsequent CPT pulses, the smaller the $Q^M_1$ is.
Hence, when the number $N$ or the duration $\tau$ of the subsequent CPT pulses is large, the duration of the preparation pulse $\tau_1$ has little impact on the spectra.
As shown in Fig.~\ref{Fig:multi-pulse}~(d) and (e), the preparation pulse has little influence on the spectra of multi-pulse CPT-Ramsey interferometry when the pulse number $N$ is large.

\subsubsection{Periodic pulse sequence}

\begin{figure}[ht!]
  \includegraphics{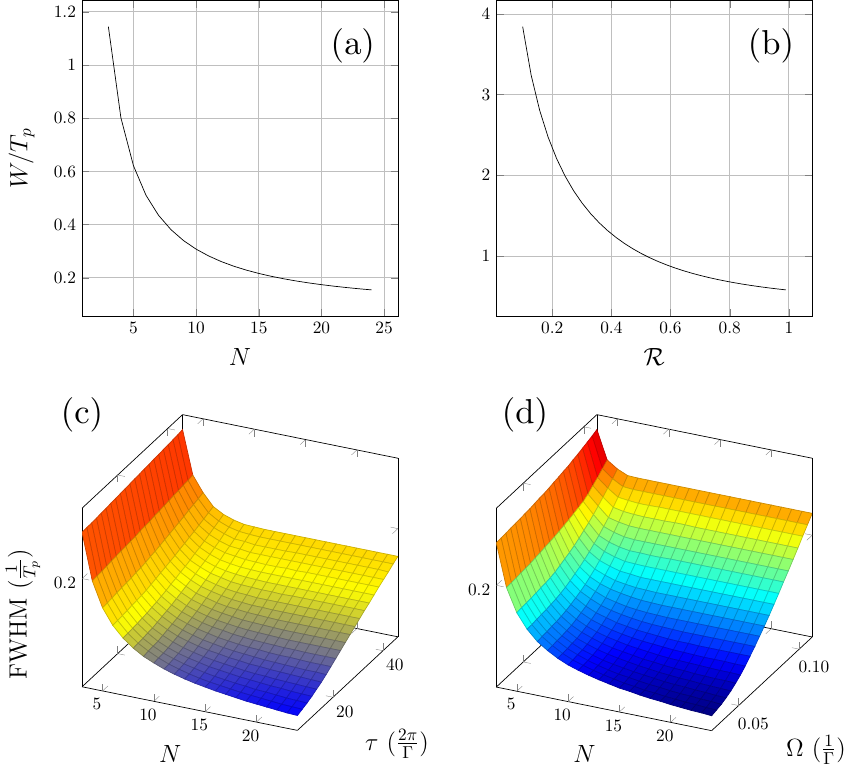}
  \caption{\textbf{The linewidth of multi-pulse CPT-Ramsey spectrum.} (a) The linewidth $W$ versus the pulse number $N$.
  (b) The linewidth $W$ versus the reflection coefficient $\mathcal{R}$.
  (c) The numerical results of FWHM versus the pulse number $N$ and the pulse duration $\tau$ with the Rabi frequency $\Omega=0.04\Gamma$.
  (d) The numerical results of FWHM versus the pulse number $N$ and the Rabi frequency $\Omega$ with the pulse duration $\tau=10\times\frac{2\pi}{\Gamma}$.
  %
  }
  \label{Fig:LineShape}
\end{figure}

With a large amount of pulse number $N$, the influence of the first CPT pulse duration can be neglected.
For simplicity, many experiments use periodic multi-pulse sequences~\cite {Yun_2012}.
Under a periodic CPT pulse sequence with interval $T$, pulse number $N$, pulse duration $\tau$, and Rabi frequency$\Omega$, according to Eq.~\eqref{Eq:analytical}, we have
\begin{equation}
  \rho_{12}(\delta)=f(\delta)\sum_{l=0}^{N-1}\mathcal{R}^{l}
  \exp\left(-\mathrm{i}l\delta T\right)\mathcal{T}.
  \label{Eq:FP}
\end{equation}
This is a temporally analog to light passing through the Fabry-P\'erot resonator~\cite{fang_temporal_2021}, in which $\mathcal{R}\equiv \exp\left(-\frac{\Omega^2}{\Gamma}\tau\right)$ takes the role the reflection coefficient and $\mathcal{T}\equiv 1-\exp\left(-\left(\frac{\Omega^2}{\Gamma}+\mathrm{i}\delta\right)\tau\right)$ corresponds to the transmission coefficient.
Using the series summation, Eq.~\eqref{Eq:FP} can be simplified as
\begin{equation}
  \rho_{12}(\delta) = f(\delta)\mathcal{T}S,
  \label{Eq:series_sum}
\end{equation}
with
\begin{equation}
  S=\frac{1-R^{N}\exp(-\mathrm{i}N\delta T)}{1-R\exp(-\mathrm{i}\delta T)}.
\end{equation}
For a multi-pulse CPT-Ramsey interferometry, $f(\delta)$ and $\mathcal{T}$ are flatter than $S$ near the resonance point.
The lineshape of $\rho_{12}$ is mainly described by $S$, as shown in Fig.~\ref{Fig:LineShape}.
If the pulse number $N\to \infty$, the full width at half maximum (FWHM) $2\pi \Delta\nu_{\infty}$ of $\Re{\rho_{12}}$ is determined by the reflection coefficient $\mathcal{R}$ and the pulse period $T_p=T+\tau$, which is the Airy distribution~\cite{ismail_fabry-perot_2016},
\begin{equation}
  2\pi \Delta \nu_\infty = \frac{4}{T_p}\arcsin\left(\sqrt{\frac{\left(1-\mathcal{R}\right)^2}{2\left(\mathcal{R}^2+1\right)}}\right).
  \label{Eq:nuinfty}
\end{equation}
Eq.~\eqref{Eq:nuinfty} is valid in the saturation region of $\mathcal{R}^N \ll 1$.

For a finite pulse number, the FWHM of Eq.~\eqref{Eq:series_sum} cannot be exactly given~\cite{pissadakis_bragg_2000}.
However, we can calculate the Lorenztian linewidth through the Taylor expansion for the $\frac{1}{\mathrm{Re}(S)}$,
\begin{equation}
  \frac{1}{\mathrm{Re}\left(S\right)} = F_0 + \frac{1}{2}F_2 T_p^2 \delta^2 + \mathcal{O}\left(T^4\delta^4\right),
\end{equation}
with
\begin{equation}
  F_0 = \frac{1-\mathcal{R}}{1-\mathcal{R}^N},
  \label{Eq:F0}
\end{equation}
and
\begin{equation}
  F_2 = \frac{\mathcal{R}\left(1+\mathcal{R}\right)}{\left(1-\mathcal{R}\right)\left(1-\mathcal{R}^N\right)} - \frac{N^{2} \mathcal{\mathcal{R}}^{N} \left(1 - \mathcal{\mathcal{R}}\right) + 2 N \mathcal{R}^{N + 1}}{\left(1 - \mathcal{R}^{N}\right)^{2}}.
  \label{Eq:F2}
\end{equation}
%
%
The real part of $S$ can be approximated as the Lorentzian form
\begin{equation}
  \mathrm{Re}(S)\approx\widetilde{S} = A\frac{\left(\frac{1}{2}W\right)^2}{\left(\frac{1}{2}W\right)^2+\delta^2}.
\end{equation}
Here, $A=\frac{1}{F_0}$ and $W=\frac{2\sqrt{2}}{T}\sqrt{\frac{F_0}{F_2}}$ are the amplitude and FWHM of $\widetilde{S}$.
Substituting Eq.~\eqref{Eq:F0} and Eq.~\eqref{Eq:F2}, we get that
\begin{equation}
  \begin{aligned}
    W = \frac{2}{T_p}\left[\frac{2\left(\mathcal{R}^2+\mathcal{R}\right)}{\left(1-\mathcal{R}\right)^2} - \frac{2N^2\mathcal{R}^N}{1-\mathcal{R}^N} -\frac{4N\mathcal{R}^{N+1}}{\left(1-\mathcal{R}^N\right)\left(1-\mathcal{R}\right)}\right]^{-\frac{1}{2}}.
  \end{aligned}
  \label{Eq:W}
\end{equation}

In Fig.~\ref{Fig:LineShape}~(a) we show $W$ versus the pulse number $N$, where $W$ decreases with the pulse number $N$.
Intuitively, more pulses will lead to a narrower linewidth.
Since $\mathcal{R}$ decreases with both $\tau$ and $\Omega$, larger $\Omega$ and $\tau$ will result in a larger linewidth.
Larger values of $\Omega$ and $\tau$ mean fewer pulses needed to reach saturation and fewer pulses contributing to multi-pulse interference, therefore the linewidth becomes broader.
For illustration, Fig.~\ref{Fig:LineShape}~(c) shows the change of FWHM with the pulse number $N$ and the pulse duration $\tau$ when $\Omega=0.04\Gamma$.
While Fig.~\ref{Fig:LineShape}~(d) displays the change of FWHM versus the pulse number $N$ and the Rabi frequency $\Omega$ when $\tau=10\times\frac{2\pi}{\Gamma}$.

\subsubsection{Anti-symmetric multi-pulse CPT-Ramsey spectroscopy with frequency shift}
\begin{figure}
  \includegraphics{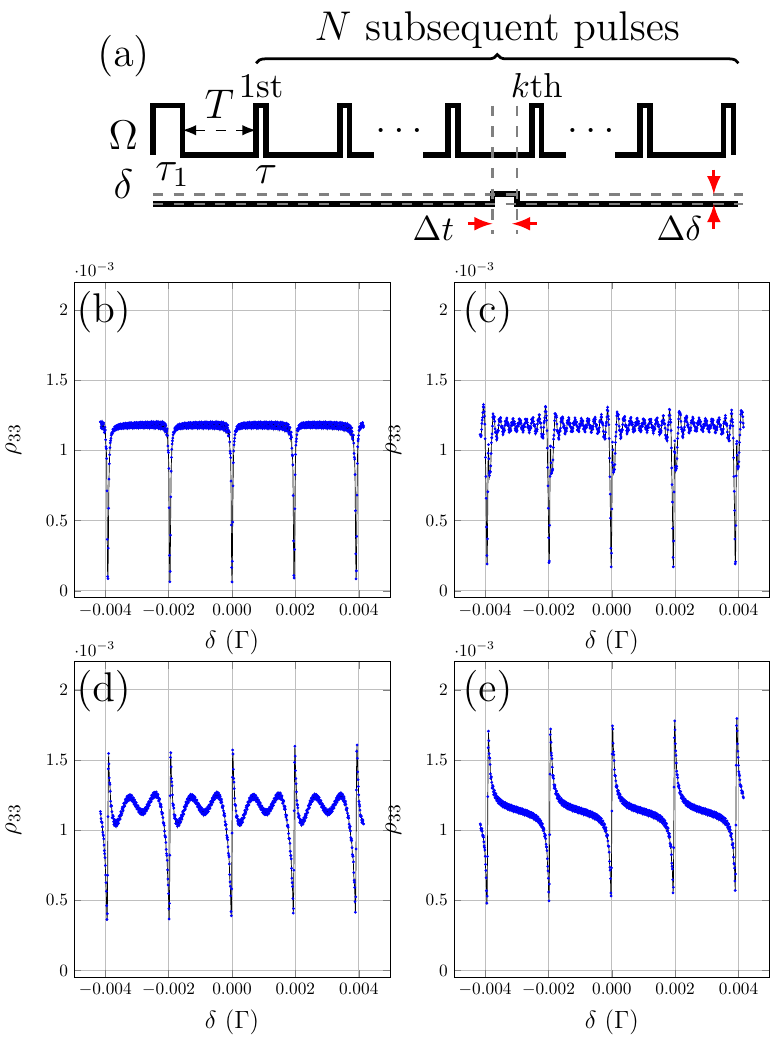}
  \caption{\textbf{Multi-pulse CPT-Ramsey interferometry with frequency shift.}
  (a) The timing sequence of Rabi frequency $\Omega$ with a frequency shift applied right before the $k$-th subsequent pulse.
  $\tau_1=200\times\frac{2\pi}{\Gamma}$ is the preparation time, $T=500\times\frac{2\pi}{\Gamma}$ is the dark time, $\tau=10\times\frac{2\pi}{\Gamma}$ and $N=40$ are respectively the duration and number of the applied periodic pulses.
  $\Delta \nu=0.0025\Gamma$ and $\Delta t=100\times\frac{2\pi}{\Gamma}$ are the magnitude and duration of frequency shift.
  The multi-pulse CPT-Ramsey spectra with the frequency shift happen
  (b) right after the preparation pulse $\tau_1$,
  (c) right before the last $10$-th pulse,
  (d) right before the last third pulse,
  (e) right before the last pulse.}
  \label{Fig:multi-pulseFrequency}
\end{figure}
As mentioned in Sec.\ref{Sec:frequencyJump}, one may prefer to use an anti-symmetric spectrum for frequency locking.
In a multi-pulse CPT-Ramsey interferometry, applying a frequency shift $\Delta\delta$ for a duration of $\Delta t$ before the $k$-th subsequent pulse [as illustrated in Fig.~\ref{Fig:multi-pulseFrequency}~(a)], or introducing a phase jump $\Delta \Phi$, can alter the spectrum horizontal symmetry to be anti-symmetric.
However, how to determine pulse sequence with frequency shift or phase jump is still challenging.
Here, we use our analytical formula to address this issue.

Taking the frequency shift and the phase jump before the $k$-th subsequent pulse, we can divide the interference term into three parts
\begin{equation}
  Q=Q^{M\prime}_1+Q^{M\prime}_2+Q^{M\prime}_3,
\end{equation}
where
\begin{equation*}
  \begin{aligned}
  Q^{M\prime}_1 =& \exp\left(-\frac{\Omega^2}{\Gamma}N\tau\right)\left(1-\exp\left(-\left(\frac{\Omega^2}{\Gamma}+\mathrm{i}\delta\right)\tau_1\right)\right)\\
  &\times \exp\left(-\mathrm{i}\delta N\left(T+\tau\right)\right)\exp\left(-\mathrm{i}\Delta\delta \Delta t + \mathrm{i}\Delta\Phi\right),
  \end{aligned}
\end{equation*}
\begin{equation*}
  \begin{aligned}
    Q^{M\prime}_2 = &\sum_{i=1}^{k-1}\exp\left(-\frac{\Omega^2}{\Gamma}(N-i)\tau\right)\\
    &\times \left(1-\exp\left(-\left(\frac{\Omega^2}{\Gamma}+\mathrm{i}\delta\right)\tau\right)\right)\\
  &\times \exp\left(-\mathrm{i}\delta (N-i)\left(T+\tau\right)\right)\exp\left(-\mathrm{i}\Delta\delta \Delta t + \mathrm{i}\Delta\Phi\right),
  \end{aligned}
\end{equation*}
and
\begin{equation*}
  \begin{aligned}
    Q^{M\prime}_3 = &\sum_{i=k}^{N}\exp\left(-\frac{\Omega^2}{\Gamma}(N-i)\tau\right)\\
    &\times \left(1-\exp\left(-\left(\frac{\Omega^2}{\Gamma}+\mathrm{i}\delta\right)\tau\right)\right)\\
  &\times \exp\left(-\mathrm{i}\delta (N-i)\left(T+\tau\right)\right).
  \end{aligned}
\end{equation*}
The frequency shift $\Delta \delta$ affects the horizontal symmetry of $Q^{M\prime}_1$ and $Q^{M\prime}_2$, but $Q^{M\prime}_3$ remains horizontally symmetric.
If the frequency shift satisfies $\Delta \delta \Delta t=\pi/2$, it will adjust $Q^{M\prime}_1$ and $Q^{M\prime}_2$ from symmetric into antisymmetric in the horizontal direction.
To obtain a horizontally antisymmetric spectrum, the contribution of the horizontally symmetric term $Q^{M\prime}_3$ should be small.
Thus, it is a natural choice to apply frequency shift right before the last pulse to suppress $Q^{M\prime}_3$.
In Fig.~\ref{Fig:multi-pulseFrequency}~(b)-(e), we applied the frequency shifts with $\Delta \delta \Delta t=\pi/2$ after the preparation pulse $\tau_1$, before the last $10$-th pulse, before the last third pulse, and before the last pulse, respectively.
The preparation pulse is $\tau_1=200\times\frac{2\pi}{\Gamma}$ and the pulse interval is $T=500\times\frac{2\pi}{\Gamma}$.
The duration $\tau=10\times\frac{2\pi}{\Gamma}$ and the pulse number $N=40$.
Clearly, as the delay of frequency shifts, $Q^{M\prime}_3$ decreases and the spectrum tends to become antisymmetric.
Thus, our analytic analysis can provide a straightforward way to design the multi-pulse sequence for CPT-Ramsey interferometry, which should be beneficial for developing high-accuracy schemes such as auto-balanced Ramsey spectroscopy~\cite{abdel_hafiz_toward_2018,abdel_hafiz_symmetric_2018,yudin_generalized_2018}.

\section{Discussion}

In conclusion, starting from the Lindblad equation, we derive an analytical formula to describe the multi-pulse CPT-Ramsey interferometry with arbitrary pulse sequence.
The analytical formula can potentially optimize the pulse sequence and analyze the influence of time-dependent detuning.
We illustrate the validity of the analytical result with the popular CPT-Ramsey scenarios and obtain the sort of views.

For a two-pulse CPT-Ramsey interferometry, we study the influence of the preparation and the detection pulse.
We quantitatively show that the preparation pulse should as long as possible to gain a larger spectrum amplitude,
and the detection pulse should be small to avoid destroying the CPT coherence.
The frequency shift or phase jump will change the spectrum symmetry.
The analytical results show that a long preparation pulse and a small detection pulse are required to obtain the antisymmetric spectrum.
For multi-pulse cases, the role of preparation pulses becomes less significant as the number of subsequent pulses increases.
As the number of pulses increases, the side peaks are continuously destroyed by interference, and we obtain a high-contrast central peak.
By adding a frequency shift or phase jump right before the last pulse, we obtain an antisymmetric multi-pulse CPT-Ramsey spectrum.
Our theoretical results can be applied to design novel multi-pulse and frequency modulated CPT-Ramsey schemes.
Under the condition of the period pulse sequence, we find out the approximate Lorenztian line shape of the spectrum and get the relationship between FWHM and the parameters of the CPT pulse sequence.
It quantitatively grapes the role of multi-pulse interference.

For the multi-pulse and frequency regulation CPT-Ramsey interferometry, there are many potential applications such as  CPT clock and CPT magnetometers.
Effective optimization methods are conducive to efficiently improving the measurement accuracy.
In this work, we eliminate the one-photon detunings and the effect of light shift is ignored.
However, the real systems always contain many states, deriving light shift from a three-level system is not meaningful.
As our analytical results can describe the situations of time-dependent detuning, we can also analyze the impact of CPT pulses on the present light shift by adding the light shift into detuning.
These are deserved for further investigations.

\begin{acknowledgments}
This work is supported by the National Key Research and Development Program of China (Grant No. 2022YFA1404104), the National Natural Science Foundation of China (Grant No. 12025509), and the Key-Area Research and Development Program of GuangDong Province (Grant No. 2019B030330001).
\end{acknowledgments}


%

\end{document}